\documentclass[final,3p,times]{elsarticle}

\usepackage{amssymb}
\usepackage{amsmath}
\journal{Journal of Subatomic Particles and Cosmology}
\begin{document}
\begin{frontmatter}

\title{Extracting the QCD speed of sound in the presence of quantum fluctuations}

\author[A]{Yu-Shan Mu}
\ead{22110190043@m.fudan.edu.cn}
\author[B,C]{Li Yan}
\ead{cliyan@fudan.edu.cn}
\author[A,C,D]{Xu-Guang Huang}
\ead{huangxuguang@fudan.edu.cn}
\affiliation[A]{organization={Physics Department and Center for Particle Physics and Field Theory, Fudan University},
             city={Shanghai},
             postcode={200438},
             country={China}}

\affiliation[B]{organization={Institute of Modern Physics, Fudan University},
             city={Shanghai},
             postcode={200438},
             country={China}}

\affiliation[C]{organization={Key Laboratory of Nuclear Physics and Ion-beam Application (MOE), Fudan University},
             city={Shanghai},
             postcode={200438},
             country={China}}

\affiliation[D]{organization={Shanghai Research Center for Theoretical Nuclear Physics, National Natural Science Foundation of China and Fudan University},
             city={Shanghai},
             postcode={200438},
             country={China}}

\begin{abstract}
The equation of state of strongly interacting matter plays a 
central role in characterizing
quark--gluon plasma created in relativistic heavy-ion collisions. It has recently been recognized that the QCD speed of sound 
is responsible for
the linear correlations between the mean transverse momentum and charged-particle multiplicity in ultra-central collisions. In realistic collisions, however, this thermodynamic correlation 
is inevitably biased by quantum-initiated event-by-event fluctuations. In this work, we present a novel statistical framework to correct this bias by 
leveraging the Gaussianity condition of the quantum fluctuations. Furthermore, we demonstrate that this approach can also serve as a probe of thermalization. Validated by the event-by-event hydrodynamic simulations, the extracted speed of sound in our framework is successfully consistent with the predictions from lattice QCD, from large to small collision systems.
\end{abstract}


\end{frontmatter}

\section{Introduction}

Relativistic heavy-ion collisions provide a unique opportunity to investigate the properties of strongly interacting matter under extreme conditions. Extensive experimental measurements have revealed a variety of collective phenomena, which are successfully described by relativistic hydrodynamics~\cite{Romatschke:2017ejr,Ollitrault:2007du,Shen:2020mgh,Yan:2017ivm}. The success of the hydrodynamic description indicates that the matter created in heavy-ion collisions approaches local thermal equilibrium rapidly and behaves as 
fluid. In such a description, the equation of state (EoS), as the input of hydrodynamics, plays a central role as it determines the thermodynamic properties of the medium and controls its collective evolution~\cite{Blaizot:1987cc}.

A fundamental quantity characterizing the EoS is the speed of sound,
\begin{equation}
c_s^2\equiv\frac{\partial P}{\partial e}=\frac{\partial \ln T}{\partial \ln s},
\end{equation}
which describes the response of the pressure to variations of energy density. The temperature dependence of the speed of sound reflects the change of the underlying degrees of freedom and provides important information about the QCD phase structure. 
Extracting speed of sound from experimental observables provides a direct connection between the observed collective dynamics and the thermodynamic properties of QCD matter.

Traditionally, constraints on EoS have been obtained by comparing hydrodynamic calculations with collective observables such as anisotropic flow and particle spectra~\cite{Monnai:2017cbv}. These analyses, however, involve the interplay between 
EoS and other dynamical ingredients, including 
initial conditions and transport coefficients. A more direct access to the thermodynamic response is through the event-by-event fluctuations in ultra-central collisions~\cite{Gardim:2019xjs,Gardim:2019brr,Gardim:2024zvi,Alqahtani:2026edr}. In these collision events, the system volume saturates and 
geometrical fluctuations are strongly suppressed. Consequently, fluctuations of 
initial density profile dominate 
event-by-event variations, providing a clean connection between entropy fluctuations and the temperature response.

Based on event-by-event hydrodynamic simulations, it has been 
demonstrated that the correlation between the mean transverse momentum $[p_T]$ and the charged-particle multiplicity $N_{\rm ch}$ in ultra-central collisions is 
sensitive to the QCD speed of sound. The underlying physical picture relies on the approximate proportionality between the final-state charged-particle multiplicity and the total entropy, $N_{\rm ch}\propto S$, as well as between 
mean transverse momentum and an effective temperature, $[p_T]\propto T_{\rm eff}$. In ultra-central collisions, the system volume is nearly fixed, allowing entropy fluctuations to be interpreted mainly as fluctuations of the entropy density. Therefore, events with larger entropy correspond to a denser medium with higher temperature, resulting in a larger mean transverse momentum. The relative response between entropy and temperature is then connected to the 
speed of sound through
\begin{equation}
\label{eq:linear}
\frac{\Delta_p}{[p_T]_0}
= c_s^2 \frac{\Delta_N}{N_0},
\end{equation}
where $[p_T]_0$ and $N_0$ denote the event-averaged values of the mean
transverse momentum $[p_T]$ and multiplicity $N_{\rm ch}$, respectively, and
$\Delta_p\equiv [p_T]-[p_T]_0$ and
$\Delta_N\equiv N_{\rm ch}-N_0$ represent their event-by-event variations.
This approach has recently been explored experimentally, providing a new possibility to constrain the QCD equation of state from heavy-ion collision data~\cite{CMS:2025mbm,CMS:2024sgx,ATLAS:2024jvf,ALICE:2025rtg}.

Neverthelss, the extraction of the speed of sound from this correlation requires a careful treatment of quantum fluctuations. In realistic collisions, the observed fluctuations are not generated solely by the thermodynamic response. Quantum fluctuations originating from the initial state and the particle production process 
inevitably present and 
propagate through the dynamical evolution. They broaden the correlation between $[p_T]$ and $N_{\rm ch}$ and 
introduce a systematic bias in the extracted response coefficient~\cite{Mu:2025gtr}.

In this work, we investigate the role of quantum fluctuations in the extraction of the QCD speed of sound. We introduce a statistical framework that separates the thermodynamic response from the quantum fluctuation contribution. For a locally thermalized system, the quantum fluctuation component is expected to follow a Gaussian distribution, allowing the physical speed of sound to be determined through a Gaussianity condition. Furthermore, deviations from Gaussianity provide a sensitive probe of thermalization in relativistic heavy-ion collisions.

\section{Quantum fluctuations and extraction of the speed of sound}

The relation between the mean transverse momentum $[p_T]$ and charged-particle multiplicity $N_{\rm ch}$ provides a direct connection between experimentally measurable 
and the thermodynamic response of the medium. However, the relation ~\eqref{eq:linear} only describes the deterministic thermal response of the system. In realistic heavy-ion collisions, event-by-event fluctuations are inevitably present. These fluctuations originate from the quantum nature of the initial state, including fluctuations of the position of nucleons, the partonic structure of the nucleons, and the particle production process. They are subsequently propagated through the hydrodynamic evolution and contribute to the final state observables.

\begin{figure*}[t]
\begin{center}
\includegraphics[width=.90\textwidth] {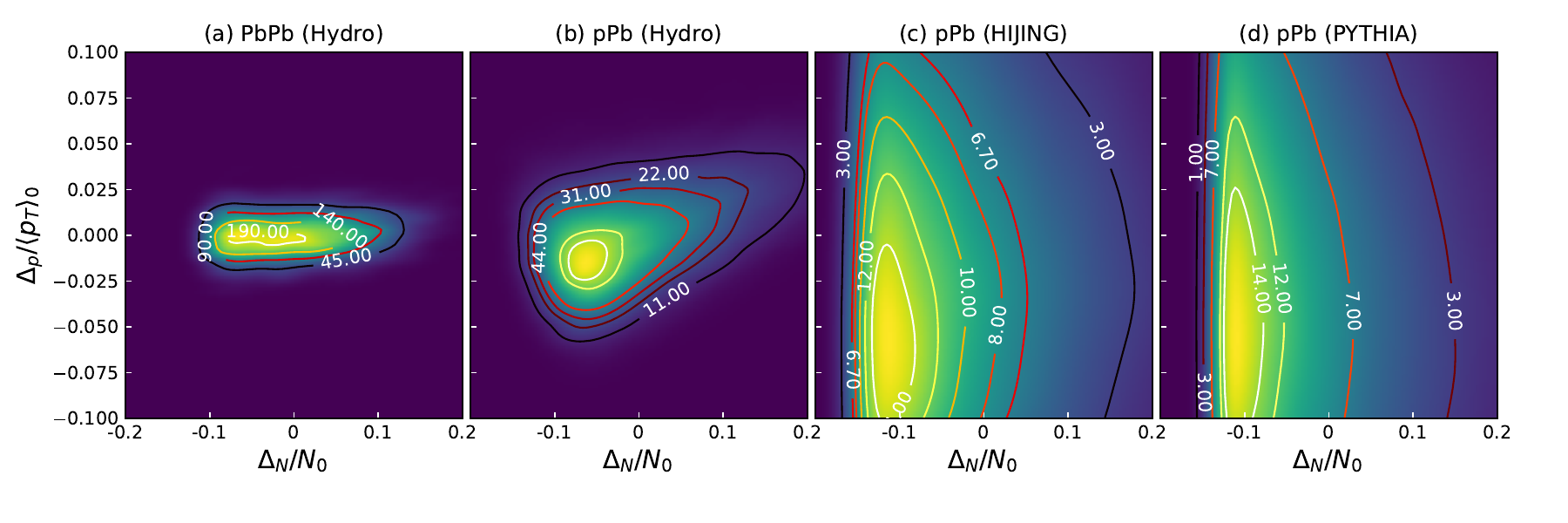}
\caption{
\label{fig:2d_dis} The normalized two-dimensional joint probability distribution of $\mathcal{P}(\Delta_N,\Delta_p)$, for the central Pb-Pb (0-5\%) and p-Pb (0-2\%) collisions at $\sqrt{s_{NN}}= 5.02$ TeV, from simulations using hydrodynamics, HIJING and PYTHIA. The tilted tip of the distributions (except PYTHIA) implies a positive correlation between $\Delta_p$ and $\Delta_N$.
}
\end{center}
\end{figure*}

As a consequence, the measured correlation between $\Delta_p$ and $\Delta_N$ does not correspond to a single response line but rather forms a two-dimensional distribution with finite width, as shown in Fig.~\ref{fig:2d_dis}. A direct linear fit to this distribution therefore mixes the thermodynamic response with the quantum fluctuation contribution and introduces a systematic bias in the extracted speed of sound.

To separate these two effects, we generalize the previous relation with an extra stochastic contribution $\delta$,
\begin{equation}
\frac{\Delta_p}{[p_T]_0} = c_s^2 \frac{\Delta_N+\delta}{N_0},
\label{eq:stochastic}
\end{equation}
The 
thermodynamic response of the thermalized medium 
is independent of microscopic quantum fluctuations. Therefore, the statistical properties of $\delta$ provide an additional constraint on extracting the physical value of $c_s^2$.

For a locally thermalized system, the thermodynamic response can be treated as independent of quantum fluctuations, and the fluctuation component $\delta$ is expected to approach a Gaussian distribution according to the central limit theorem~\cite{Samanta:2023amp,Samanta:2023kfk}. We therefore determine the physical value of $c_s^2$ by requiring the skewness of the fluctuation distribution to vanish,
\begin{equation}
\{\delta^3\}_c=0 .
\end{equation}
This condition leads to a cubic equation for $c_s^2$,
\begin{equation}
\label{eq:gauss}
(c_s^2)^3 \frac{\{\Delta_N^3\}}{N_0^3} 
- 3 (c_s^2)^2 \frac{\{\Delta_N^2 \Delta_p\}}{N_0^2 [p_T]_0}
+ 3 c_s^2 \frac{\{\Delta_N \Delta_p^2\}}{N_0 [p_T]_0^2}
- \frac{\{\Delta_p^3\}}{[p_T]_0^3}=0 .
\end{equation}
All coefficients in this equation are determined entirely by the experimentally accessible moments of the joint distribution of $\Delta_p$ and $\Delta_N$. Solving this cubic equation gives the speed of sound with the effect of quantum fluctuations explicitly taken into account. Event-by-event hydrodynamic simulations demonstrate that the extracted values of $c_s^2$ using this method are in good agreement with lattice-QCD predictions~\cite{Mu:2025gtr,HotQCD:2014kol}, validating the applicability of the Gaussianity-based approach for studying the thermodynamic response of the QCD medium.

\section{Quantum fluctuations as a probe of thermalization}

The statistical properties of the quantum fluctuation component provide 
not only 
a method to extract the speed of sound, but also 
captures thermalization achieved in heavy-ion collisions. The underlying idea is that local thermalization establishes a separation between the macroscopic hydrodynamic response and microscopic fluctuations. Once the system reaches local equilibrium, the thermal response is dominant, while the residual fluctuations behave as approximately independent Gaussian noise.

\begin{figure*}[t]
\begin{center}
\includegraphics[width=1.0\textwidth] {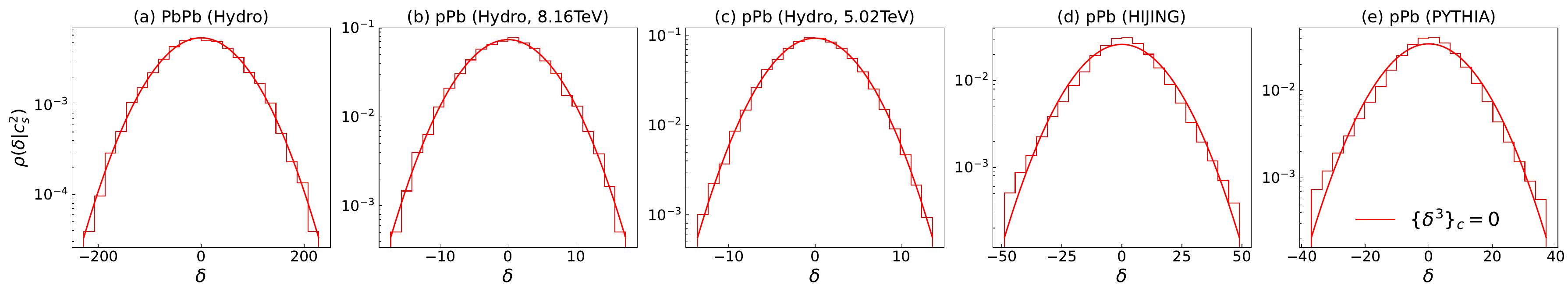}
\caption{
\label{fig:gauss} The normalized probability distribution $\rho(\delta|c_s^2)$,  with $c_s^2$ determined by the zero skewness condition $\{\delta^3\}_c=0$ (red solid histogram). 
For comparison, the Gaussian distributions with zero mean and the variance $\{\delta^2\}^{1/2}$ determined with respect to the extracted $c_s^2$ values are shown by red solid line.
}
\end{center}
\end{figure*}

Therefore, the Gaussianity of $\delta$ can also serve as a signature of thermalization. As shown in Fig.~\ref{fig:gauss}, in hydrodynamic simulations, $\delta$ exhibits a nearly Gaussian distribution, and the extracted speed of sound remains consistent with the thermodynamic expectation. This indicates that, in a locally thermalized system, the thermodynamic response and quantum fluctuations can be effectively separated, and the observed $[p_T]$--$N_{\rm ch}$ correlation originates from the collective response of the medium.

In contrast, without thermalization, the response between mean transverse momentum and multiplicity is no longer governed by a thermodynamic relation. 
The separation between the thermodynamic response and quantum fluctuations breaks down, and $\delta$ no longer exhibits the Gaussian behavior. In this case, neither a physical extracted value of the speed of sound, nor Gaussianity of $\delta$ is expected, indicating that the system does not exhibit the thermodynamic behavior of an equilibrated medium.

The deviation from Gaussianity can be quantified by higher-order cumulants of the fluctuation distribution. For example, the standardized kurtosis,  
\begin{equation}
\kappa_4 = \frac{\{\delta^4\}}{\{\delta^2\}^2}-3 ,
\end{equation}
vanishes for an ideal Gaussian distribution. Therefore, $\kappa_4$ provides a quantitative measure of the deviation from Gaussian fluctuations. A value close to zero indicates that the fluctuation component is consistent with the expectation of a locally thermalized system, where quantum fluctuations can be separated from the thermodynamic response. In contrast, a significant deviation from zero indicates the breakdown of this separation and the absence of a purely Gaussian quantum fluctuation component.

\section{Summary}

In this work, we have investigated the impact of quantum fluctuations on the extraction of the QCD speed of sound from the event-by-event correlation between the mean transverse momentum and charged-particle multiplicity in ultra-central heavy-ion collisions. While this correlation provides a direct connection between final-state observables and the thermodynamic properties of the medium, quantum fluctuations introduce additional bias that can modify the extracted response coefficient.

We have developed a statistical framework that separates the thermodynamic response from the quantum fluctuation contribution. For a locally thermalized system, the fluctuation component is approximately Gaussian, and the physical speed of sound can be determined by imposing the Gaussianity condition. Compared with the conventional linear-fit method, this approach avoids the bias caused by the quantum fluctuations and provides a reliable extraction of the equation-of-state information.

Furthermore, we have shown that the statistical properties of quantum fluctuations provide a sensitive probe of thermalization. Hydrodynamic simulations exhibit nearly Gaussian fluctuations and yield physically consistent values of the speed of sound, whereas non-thermal transport models generate significant non-Gaussian fluctuations and extract unphysical values. These results demonstrate that quantum fluctuations are not merely a source of uncertainty, but contain valuable information about the dynamical evolution and thermalization of the strongly interacting matter created in heavy-ion collisions.

\bibliographystyle{elsarticle-num}
\bibliography{ref}

\end{document}